\documentclass[prb,twocolumn,amsmath,amssymb]{revtex4}
\begin{document}
\author{Kevin Leung and Susan B.~Rempe}
\affiliation{Sandia National Laboratories, MS 1415 \& 0310,
Albuquerque, NM 87185}
\date{\today}
\title{{\it Ab initio} Molecular Dynamics Study of Glycine Intramolecular
Proton Transfer in Water}

\input epsf

\begin{abstract}

We use {\it ab initio} molecular dynamics simulations
to quantify structural and thermodynamic properties of a model
proton transfer reaction that converts a neutral glycine molecule,
stable in the gas phase, to the zwitterion that predominates in
aqueous solution.  We compute the potential of mean force associated
with the direct intramolecular proton transfer event in glycine.
Structural analyses show that the average hydration number (N$_{\rm w}$)
of glycine is not constant along the reaction coordinate,
but rather progresses from N$_{\rm w}$=5 in the neutral molecule to
N$_{\rm w}$=8 for the zwitterion.  We report
the free energy difference between the neutral and charged glycine
molecules, and the free energy barrier to proton transfer. Finally, we identify
approximations inherent in our method and estimate corresponding corrections
to our reported thermodynamic predictions.

\end{abstract}

\maketitle



\section{Introduction}

Proton transfer plays diverse but fundamentally important roles in
aqueous solution phase chemistry.  Transfer reactions in water, aided 
by the indistinguishable nature of protons, produce highly mobile
charge carriers utilized in fuel cells\cite{paddison} and biological systems
\cite{decoursey} for
electrical energy conduction.  Protonation and deprotonation reactions
on ionizable substrates activate and modulate biochemical reactions.
For example, in Rubisco, the enzyme that catalyzes the first reaction in 
carbon fixation, many steps in the reaction mechanism consist of 
protonation or deprotonation events on the enzyme-substrate
complex.\cite{andersson}  
In biological ion channels, changes in
protonation states along the ion conduction path modify 
transport function,\cite{miller1} and changes in protonation state in the gating
region of some channels presumably trigger opening or closing 
of the channels.\cite{miller2}  
In order to understand such chemical processes associated with proton transfer
in molecular detail, it is crucial to have methods that yield
accurate estimates of the accompanying free energy changes. 

Free energies, which characterize molecular interactions and
govern the likelihood and rates of chemical reactions and
conformational changes, are among the most important properties
to be computed in aqueous environments.
Since its introduction, the {\it ab initio} molecular dynamics 
approach (AIMD) has
demonstrated remarkable success in predicting and modeling the
hydration structure, vibrational signature, and electronic
properties of ions and molecules in water.\cite{aimd_review}  In
recent years, it has also found increasing utility in computing
thermodynamic properties,\cite{siepmann} including free energy
differences by potential of mean force methods.\cite{klein,sprik}
These methods have successfully reproduced several pKa's for
deprotonation reactions in water.

In fact, 
the recent introduction
of various methods to calculate 
free energies,
such as reaction-path finding techniques,\cite{autoion}
targeted molecular dynamics\cite{tmd} and related methods,
and a new theory of the liquid state that emphasizes the local
hydration environment,\cite{pratt_pct}
has created the potential for combining AIMD and unbiased, rigorous
predictions of free energy barriers and differences.  
Relatively few 
aqueous phase free energy
calculations exist in the literature based explicitly on AIMD data.  
To benchmark new
advances in free energy methods, 
AIMD free energy calculations
on simple, representative systems (especially proton transfer reactions)
would be extremely valuable.

A convenient and popular model system for studying 
proton transfer events in aqueous
environments is the intramolecular proton transfer of glycine in water.
Water triggers zwitterion formation from the neutral molecule stable
in the gas phase.
Since the stability of the zwitterion relative to neutral conformers
in water is entirely due to glycine-water
interactions,\cite{jensen} theoretical predictions
for glycine in water are sensitive to the parameterization of
water-glycine functional group interactions.  This
makes the glycine intramolecular proton transfer reaction a sensitive
and valuable benchmark for calculating hydration effects on
biological functional groups.  
In fact, when new methods are devised to sample molecular conformations
at finite temperature quantum mechanically, glycine is
often a model of choice as a first application.\cite{blochl,bandy}

Part of the reason for the popularity of glycine as a model
for proton transfer studies in proteins
is that its tautomerization free energy difference
and barriers have been
measured.\cite{sheinblatt,wada,chang,grunwald,slifkin,haberfield,kameda3}
Glycine can exist in many neutral form conformers in addition
to the zwitterion tautomer, as illustrated in Fig.~1.
In the gas phase, the zwitterion (ZW) is
{\it unstable} and spontaneously collapses to the neutral form (NF) conformer
``IIp.''\cite{conform,ding}  In contrast, the ZW is more stable than any
neutral form in water due to large local charges and a large dipole moment.
A free energy barrier of 14.3 kcal/mol is associated with the ZW$\rightarrow$NF
interconversion,\cite{slifkin} and ZW is more stable
by 7.27 kcal/mol.\cite{wada}  From these results, the reverse,
NF$\rightarrow$ZW reaction is estimated to have a free energy
barrier of $\sim 7$ kcal/mol and there are indications that the
ZW$\rightarrow$NF free energy barrier is mainly entropic in
origin.\cite{slifkin} Other experimental data provides additional
information on glycine hydration.
Neutron scattering results on the hydration structure of the ammonium
group in concentrated glycine solutions have been
reported,\cite{kameda3} as well as mass and size-selected
photoelectron spectroscopic studies of hydrated glycine anions.\cite{bowen}

\begin{figure}
\centerline{\hbox{\epsfxsize=3.50in \epsfbox{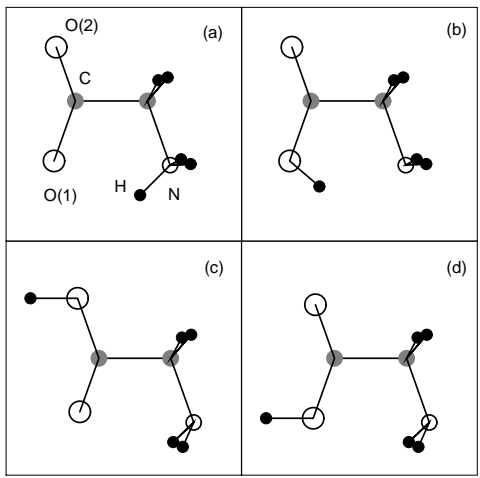}}}
\caption[]
{\label{fig1} \noindent
Glycine tautomers.  Black circles: H; grey: C; small and large open circles:
N and O, respectively.
Panel (a): zwitterion (ZW) glycine; (b)-(d) three stable/metastable
neutral form (NF) glycine molecules with configurations optimized in
the gas phase, respectively, IIp, Ip, and IIIp.\cite{conform}  Note that
some works in the literature label O(1) and O(2) in the reverse way.
}
\end{figure}

%

While experimental data makes glycine a choice system
for study, comparison of theoretical and
experimental studies of the proton transfer
between the ZW and NF tautomers is complicated by
the fact that the most stable aqueous NF conformer has not been
identified.\cite{review}
The various NF conformers are related
by rotation of the C~-~C, C~-~O, and C~-~N bonds.
Conformer Ip (Fig.~1(c))
is the most stable structure in the gas phase while IIp is $\sim 1$
kcal/mol higher in energy.\cite{sirois,conform}  
Intuitively, a direct intramolecular proton transfer from ZW to NF in water
should go first through IIp, which is closest in structure to ZW among
the neutral form conformers.   This is the proton transfer reaction that is
most accessible to finite temperature, quantum mechanic (or quantum
mechanics-based) simulations.  Thus practitioners using a
molecular dynamics simulation, based on a reactive force field fitted
to Hartree-Fock data,\cite{yamabe1,yamabe2}
have claimed to reproduce the experimental
values for ZW$\rightarrow$NF free energy difference and barrier by just
considering the IIp neutral form.  
Other quantum chemistry works, however, find a small reverse IIp$\rightarrow$ZW
barrier and interpret IIp as an intermediate, not
the final or most stable NF product.\cite{bandy,tunon00,kassab}  
To further complicate interpretation of the results,
these quantum chemistry results depend on the details of
the calculations and so are not necessarily in quantitative agreement with
each other.

In this work, we do not seek 
the most favorable form of (metastable) NF glycine in aqueous solution.  
Instead, we use the {\it ab initio} molecular dynamics (AIMD)
approach
to focus on the direct IIp$\rightarrow$ZW proton transfer in water.
AIMD treats the valence electrons of all atoms quantum mechanically, using
density functional theory (DFT).  It also samples
glycine and water conformations at finite temperature via its
molecular dynamics capability.  While AIMD is computationally costly,
it has several important advantages when treating intramolecular
proton transfer in glycine.  (a) Glycine itself is treated quantum
mechanically, which is pertinent to the breaking and making of
chemical (covalent) bonds that take place during proton transfer.
(b) The number and conformations of water molecules in the first
hydration shell, and the hydrogen bonds they form with the glycine
atoms, are allowed to fluctuate and vary.  This is
significant because hydrogen bond network fluctuations have been
known to be crucial in other prototypical proton transfer reactions in
water.\cite{tuckerman02}  (c) These first hydration shell water
molecules are treated quantum mechanically.
Widely used force field models for water and the carboxylate
(-COO$^-$) functional group, which is a crucial part of ZW glycine,
predict carboxylate group hydration
structures\cite{stamato,jorg_choo,koll_gly,koll_choo,others,gao1}
that disagree with experiments.\cite{kameda1,kameda2}  This
is true even if the carboxylate group containing ion is treated
quantum mechanically and only water is treated classically\cite{cubero}
(i.e., a QM/MM treatment\cite{qmmm}).
In contrast, AIMD predictions of hydration structure are in
good agreement with experiments.\cite{choo_ab}  Likewise,
in the one case where a QM/MM estimate of the ZW/IIp free energy
difference is reported, the theoretical prediction overestimates the
experimental value by a factor of four.\cite{shoeib}  This emphasizes the
sensitivity of this free
energy difference to glycine-water interaction parameters.

The role of water conformational changes
and water-glycine hydrogen bond fluctuations on the glycine intramolecular
proton transfer reaction is central to this work.  We dynamically sample water
conformations using AIMD.  These water
configurational changes may be especially important for the
ZW$\rightarrow$IIp reaction.  This is because 
the glycine dipole moment decreases along the reaction
coordinate, leading to a reduction in the mean hydration number.  
Our work will reveal trends concerning hydration number variations and
hydrogen bond network fluctuations as the intramolecular proton transfer
proceeds.

A number of pioneering quantum
chemistry works on glycine tautomerization have used a quantum treatment
of the glycine molecule as well as a few water molecules in the
glycine first hydration shell (i.e., a ``supermolecule'' approach).
A static optimization of the supermolecule geometry is performed, and a
polarizable dielectric continuum is used to treat the outlying
water.\cite{bandy,kassab,fern}  The advantage of such methods is
their relative computational ease, which allows the sampling of
many glycine neutral form conformations.

In connection with these studies,
our AIMD prediction of the progression in the average hydration
number around each glycine atom, as the reaction coordinate varies,
may contribute to future supermolecule
studies of proton transfer reactions in water.
Working within the framework of a statistical mechanical theory
of liquids called quasi-chemical theory,\cite{pratt} for example,
accurate hydration free energies for metal ions have been computed by
forming supermolecular clusters containing the metal ion and
the full set of water molecules
in the first hydration shell and embedded in a dielectric
continuum model of bulk water.\cite{rempe,rempe2,rempe3,rempe4}
Clusters containing one glycine and 3 or 6
water molecules have been considered in recent published quantum chemistry
studies.\cite{bandy,kassab,fern}
As will be shown in this work, this is not sufficient to complete
the first hydration shell of ZW glycine.  We will make a preliminary
attempt to quantify this effect.

The paper is organized as follows.  Section~2 describes the method
used.  Our predicted potential of mean force and correlation
functions are described in Sec.~3.  Section~4
concludes the paper with further discussions.  In the appendices,
we discuss corrections to the AIMD results by estimating the
effect of zero point energy contributions and
of using different exchange-correlation functionals, and also we address
the effect of using a finite-sized simulation cell.

\section{Method}
\label{method}

We perform {\it ab initio} molecular dynamics simulations on a system with 1 glycine
and 52 H$_2$O molecules.  Finite size effects will be addressed
in Appendix B using a simulation cell with 98 H$_2$O molecules.
The Car-Parrinello Molecular Dynamics
(CPMD)\cite{cpmd} code is applied, along with the BLYP gradient
corrected exchange correlation functional\cite{blyp_corr,blyp_ex} and
Troullier-Martins pseudopotentials.\cite{troullier}  BLYP
has been shown to yield water-water
pair correlation functions $g(r)$\cite{siepmann} as well as
hydration structures of NH$_4^+$ and HCOO$^-$
ions that are in good agreement with experiments.\cite{choo_ab,nh4}
The effect of using other, perhaps more accurate, exchange correlation
functionals will be addressed in Appendix A.

The simulation box is cubic with linear
dimension 11.76~$\AA$, which corresponds to a water density of
1.00 g/cm$^{3}$ plus the experimental glycine zwitterion volume of
72~$\AA^3$.\cite{expt_vol}
The time step used is 5 a.u.~(0.121 fs), and the deuterium
mass is assumed for all protons throughout, although we
will continue to use the word ``proton.'' The temperature is
kept at T=300~K using a thermostat unless otherwise stated.

We compute the potential of mean force, $\Delta G(R)$,\cite{helmholtz}
along a reaction coordinate 
defined as the difference between
the N~-~H and O$(1)$~-~H distance:
\begin{equation}
R=R_{\rm N-H} - R_{\rm O(1)-H}. \label{eq1}
\end{equation}
O(1) is the IIp glycine oxygen atom covalently bonded to the
acidic proton, and H is the proton being directly transferred
between O(1) and the nitrogen atom.
Progression along this coordinate represents a direct proton
transfer from the ZW to the IIp form of glycine.
For the IIp neutral conformer (Fig.~1b), 
$R_{\rm O(1)-H} \sim 1$ \AA, and
N and H are not covalently bonded, making $R_{\rm N-H}$ larger
than 1.4 \AA, and hence $R>0$.  For the ZW glycine,
$R_{\rm N-H} \sim 1$ \AA, O(1)~-~H is not covalently bonded,
and $R<0$.  
We use the umbrella sampling method\cite{chandler} (with harmonic
constraints or biasing potentials) to compute the free energy
profile along this reaction coordinate.  Since one of the sampling
windows is the unconstrained glycine zwitterion, this allows us to examine
the structure of water around stable, equilibrium zwitterion glycine,
and compute the pertinent water-glycine pair correlation functions
($g(r)$) as well.\cite{note1}

The barrier height $\Delta G^*$ depends on the reaction pathway.  A
more sophisticated approach would be to use the path-sampling
method,\cite{chandler1} which in principle can sample the possible
reaction pathways weighted by the proper statistical mechanical
weight.  This is computationally expensive, however,
and seldom implemented within an AIMD setting.   For the relatively
simple system of glycine intramolecular proton transfer, we expect
our relatively simple reaction coordinate to be adequate.
This is because the reaction coordinate is similar to the one considered in
Ref.~31
-- it describes a straightforward direct proton transfer event.  Using a
simple N~-~H bond distance reaction coordinate, Kassab {\it et al}. obtain
results similar to their own gas-phase transition state finding
calculations.\cite{kassab}
Furthermore, when we initiate a glycine IIp conformation in water,
we observe that it undergoes a direct, intramolecular proton transfer to
the ZW form in less than 1~ps.  This observation corroborates the
reaction coordinate we have chosen.

While the two oxygen atoms in ZW are in principle equivalent,
they interconvert on a relatively long time scale, and
they can be treated as distinct
for the duration of our AIMD trajectories. 
In contrast, we do observe rotation around the N~-~C
bond in our simulations, which is also 
known to occur in picosecond timescales.\cite{Nrotation}
The three protons bonded to the nitrogen atom in the ZW glycine are therefore
equivalent in our simulations, and, at any time step, the relevant proton in
Eq.~\ref{eq1} is taken as the one closest to O(1).\cite{note1}
To compute the free energy profile along $R$, a harmonic penalty function
$V(R)=A_i(R-R_i)^2$ is used to constrain $R$ in six umbrella sampling
windows with a progression of target distances $R_i$, where $A_i$ ranges
from 0.7 eV to 1.2 eV/AA (see Table~\ref{table1} for details).  Window
(a) corresponds to ZW glycine, which is stable and requires no constraint.

\begin{table}\centering
\begin{tabular}{||c|c|c|c|c|c|c||} \hline
window    & a (ZW) &   b  &   c  &   d  &   e  & f (IIp) \\ \hline
$R_i$ (\AA) &  0.0   & -0.8 & -0.5 & -0.1 &  0.2 &  0.7 \\
$A_i$ (eV)  &  0.0   &  1.0 &  1.0 &  1.2 &  1.0 &  0.7 \\ \hline
\end{tabular}
\caption[]
{\label{table1} \noindent
Constraint parameters $A$ and $B$ for the umbrella sampling windows.
See Eq.~\ref{eq1}.}
\end{table}

AIMD trajectories are initiated by inserting a ZW or IIp glycine molecule into
a simulation box with 54 water molecules, which has been 
equilibrated previously using the empirical SPC force field for water.\cite{spc}
Two water molecules overlapping the glycine molecule are removed.  We first
conduct a 10 ps QM/MM molecular dynamics trajectory for ZW and IIp
glycine.\cite{vasp}  The resulting molecular
configurations are used as the starting points for the AIMD trajectory in
the ZW and IIp windows (i.e., windows (a) and (f)).  Using AIMD,
the QM/MM IIp configuration is equilibrated for 4 ps, and then
statistics are collected for the next 10 ps.  The window (f)
constraint (Table~\ref{table1}) is always applied to keep the IIp glycine
from undergoing intramolecular proton transfer.
The starting configuration of window (e) is taken
8~ps into the IIp trajectory, equilibrated at
the new constraint (Table~\ref{table1}) for 6~ps, and run for
another 10~ps.  Window (d) is spawned from window (e) 5~ps into
its trajectory, then equilibrated for 2~ps with the new constraint.
As will be discussed, in windows (d) and (e),
statistics are collected for 20~ps.

As for window (a), we first use AIMD to re-equilibrate a QM/MM ZW
configuration at T=300~K for 10~ps.  The final configuration
of this trajectory is once again re-equilibrated at T=350~K for 1~ps,
and then statistics are collected at this temperature for 10~ps.
Both windows (b) and (c) are spawned from the configuration at
the end of the window (a) trajectory, equilibrated for 1~ps with
the new constraint, and then statistics are collected for 10~ps.
These two windows are thermostated at T=350 and 300~K, respectively.
The reason for using a higher temperature, and the small effect this has
on the potential of mean force, will be discussed later on.

Our AIMD trajectories within each sampling window last 10 or 20 ps.
Our total AIMD trajectory length is similar to that used to correctly
reproduce experimental results for deprotonation of histidine
residues.\cite{klein}
Note that the reactive force field based glycine intramolecular
proton transfer work of
Ref.~28
uses trajectories only a few times longer than the one reported in our work.
Furthermore, our use of 
longer trajectories in the windows that contribute most to $\Delta G$,
and a higher temperature for the ZW window that only weakly contributes
to $\Delta G$ (see below) further reduces statistical uncertainties.
We estimate the cumulative statistical uncertainty of our predicted
$\Delta G$ and $\Delta G^*$ between IIp and ZW glycine to be of order
$\sim 1.4$ kcal/mol.
($\Delta G^*$ for the reverse reaction, IIp$\rightarrow$ZW, is
much smaller.)  As a result, one limitation of our AIMD work is that
we cannot resolve the temperature dependence of $\Delta G$ for
this intramolecular proton transfer reaction.  Nevertheless, the
potential of mean force computed using AIMD potentially can be used
to calibrate or refine force fields, which then can be applied to 
address the entropic contribution to $\Delta G$ accurately and
efficiently.

\section{{\it Ab initio} molecular dynamics results}

In this section, we first consider the hydration structures
of the glycine functional groups.   We also investigate
the time dependence of the hydration numbers to demonstrate
that the simulation time we use in each umbrella sampling
window is adequate.  Then we report the potential of mean
force and hydration structures along the ZW$\rightarrow$IIp
reaction coordinate.  Finally, we discuss our results in relation
with experiments and quantum chemistry based supermolecule calculations.

\subsection{Zwitterion hydration structure}

\begin{figure}
\centerline{\hbox{\epsfxsize=3.50in \epsfbox{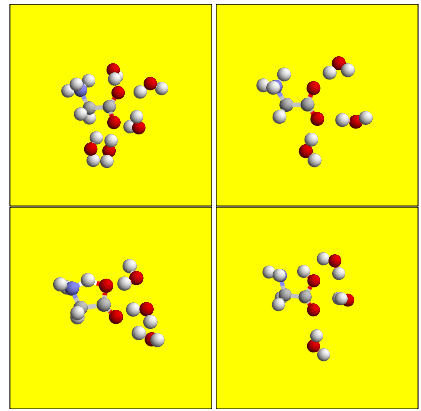}}}
\caption[]
{\label{fig2} \noindent
Sample hydration structures around just the
carboxylate/carboxylate acid group (not the amine group)
in aqueous glycine.  Only the first hydration shell is shown.
Th grey, blue, red, and white spheres represent
C, N, O, and H atoms, respectively.
Upper left panel: long lived zwitterion hydration shell configuration at
T=300 K.  Upper right: zwitterion hydration shell after a trajectory
of 5~ps at T=350 K.  Lower left: window (d) (transition state region);
lower right: window (f) (neutral form glycine).
}
\end{figure}

ZW glycine, the most stable tautomer in water,
has a large gas phase dipole moment and charge separation
along the molecular framework that leads to strong interactions with water.  
Despite this, we expect, on the basis of work with formate ion
hydration,\cite{choo_ab} that the coordination number of the
carboxylate oxygens will experience large fluctuations.\cite{h_num}
We indeed find that the combined instantaneous hydration number for these
oxygens ranges from 3 to 7, as illustrated in Fig.~3.  
At T=300~K, however,
the first hydration shell of the -COO$^-$ group exhibits relatively slow
dynamics during the first $\sim 10$~ps of the AIMD trajectory.
For instance, in a 5~ps stretch, one of the water molecules bound to
O(2) briefly leaves the hydration shell, is replaced by a second water,
and then returns, displacing the newly added water molecule.  This
long-lived ZW configuration has two water molecules hydrogen bonded
to O(1) and three to O(2) (see Fig.~2).

\begin{figure}
\centerline{\hbox{\epsfxsize=3.50in \epsfbox{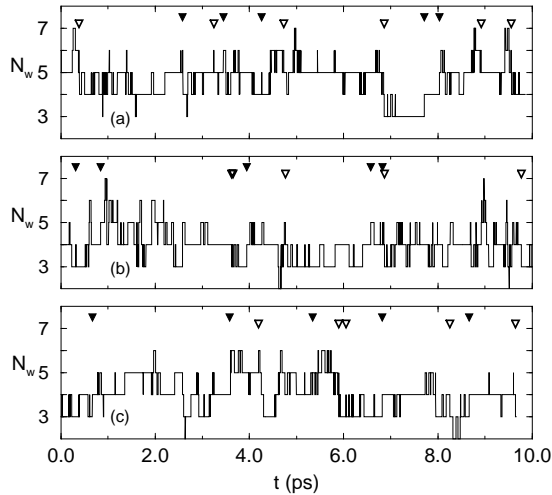}}}
\caption[]
{\label{fig3} \noindent
Time dependent hydration numbers for O(1) and O(2) combined.
(a) Window (a); (b) window (c); (c) window (d).
Based on the first minimum of the glycine-water $g(r)$, we define
a water molecule to be in the hydration shell if it one of its protons
is within 2.5 \AA\/ of the glycine oxygens. 
Filled and hollow triangles indicate where water
molecules enter the hydration shell of O(1) or O(2) for the
first time, or leave those hydration shells for at least 1 ps.
There are numerous transient fluctuations and re-entrances into
hydration shells.  For concreteness, we arbitarily define
a water molecule as ``first'' entering the hydration if it
resides there for at least 1~ps later in the trajectory, {\it and}
if it was not previously in the hydration shell for at least 2 ps.
}
\end{figure}

We did not observe similar slow dynamics in AIMD simulations of
the aqueous formate ion\cite{choo_ab} using the BLYP exchange
correlation functional, despite the fact that HCOO$^-$ is a
charged species.  We speculate that, whereas
the carbon end of the formate ion is hydrophobic, the 
-NH$^{3+}$ group in the ZW glycine forms another strong hydration
shell and concentrates water molecules in the vicinity of the
-COO$^-$ group, causing a more persistent carboxylate-water
interaction in the ZW glycine.  
Regardless of the reason, while this long-lived hydration
structure persists, the reaction coordinate $R$ will be locked
into a relatively narrow distribution.  As a result, sufficient statistics
for the ZW glycine cannot be accumulated using a 10-20 ps trajectory.
Instead, we collect statistics for windows (a) and (b) at
T=350~K.  The dynamics at this higher temperature
are much faster and involve several exchanges of water molecules between
the first hydration shell and the bulk liquid.  See Fig.~3.

As $R$ increases, the glycine dipole moment decreases,
its interaction with water weakens, and the dynamics of water molecules
around the glycine molecule become faster.  Figure~3(b) shows
that the dynamics in window
(c) and (d) are comparable to that in (a), despite the fact they
are run at T=300~K.  This indicates that an elevated temperature
is not needed for window (c) for a 10~ps trajectory.
Figure~3 actually depicts two complementary quantities, the
instantaneous hydration number of O(1) plus O(2), and the times at
which the water molecules first enter or leave the hydration shells.
The former includes all transient fluctuations and the rapid
motion due to water molecules briefly forming and breaking
hydrogen bonds with O(1) or O(2).  The latter filters out the
rapid fluctuations, but its definition is somewhat arbitary;
we require that water molecules have at least a 1~ps residence time
to be so counted.  At even higher temperatures, such that the time scale
of exchange between first hydration shells and the bulk water region
is faster than 1~ps, this criterion will have to be redefined.

\begin{figure}
\centerline{\hbox{\epsfxsize=3.50in \epsfbox{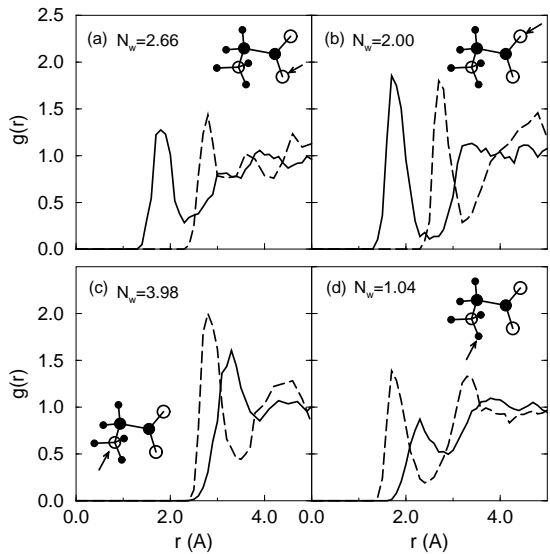}}}
\caption[]
{\label{fig4} \noindent
Pair correlation functions between four atoms in ZW glycine,
and oxygen (dashed line) and hydrogen atoms (solid line) on water
molecules.  (a) O(1): (b) O(2): (c) N; (d) one of the three
equivalent protons on the ammonium group. N$_w$ indicates the
number of water molecules around the atom, found by integrating
to the first minimum in g(r) between the solute and water oxygen atoms. 
}
\end{figure}

We note that the residence times of water molecules
in the hydration shells are sensitive
to the strength of interaction between carboxylate oxygens and water
molecules; a slight increase of just 1 kcal/mol will increase them
significantly.  Hence such quantities are expected to be sensitive to
temperature as well as to the exchange correlation functional
used.\cite{choo_ab}

Figure~4 depicts the ZW glycine-water pair correlation
functions, $g(r)$, for selected atomic sites.
The two carboxylate oxygen atoms form an average of 4.7 total
hydrogen bonds with water molecules.  O(2) exhibits a hydration number
N$_{\rm w} \sim 2.66$,
in good agreement with the hydration number of the formate
ion computed using AIMD.\cite{choo_ab} O(1) forms an average of 2.0
hydrogen bonds with water molecules.  It also forms an intramolecular
hydrogen bond with one of the ammonium group protons 40\% of the time,
assuming a hydration shell radius of 2.5~\AA.  The combined hydration
number of 4.7 is smaller than that of 5.3 predicted by 
Alagona {\it et al.}\cite{koll_gly} using empirical force fields, and
is much smaller than the hydration number of 7 for carboxylate groups 
predicted by Jorgensen and Gao's OPLS force fields.\cite{gao1} 
Integrating $g(r)$ between the ammonium protons and water oxygen
atoms up to its first minimum, the ammonium group is found to have
3.0 water molecules in its first hydration shell.  
Using instead the g(r) between
the nitrogen and water oxygen, we find N$_{\rm w} \approx 4$
under its first peak, with an average N~-~O distance of 2.75 \AA.
A neutron scattering study for 5 mol~\% glycine solution deduces a similar
mean N~-~O distance of $2.85 \pm 0.05$ and
$N_{\rm w} = 3.0 \pm 0.6$, based on analysis
of the N~-~O correlation.\cite{kameda3}  
Since the ammonium protons are directly hydrogen bonded to the
water oxygens while the nitrogen atom is not, we deem the former
more useful in determining the total number of hydrogen bonds,
particularly because of the elevated temperature used in our
simulations.  Hence we estimate that AIMD predicts 3.0 hydrogen bonds
between the ammonium group of ZW glycine and water molecules.

Unlike the carboxylate oxygens, the
combined coordination number of the three protons in the -NH$_3^+$
do not exhibit large fluctuations, and thus these fluctuations should
not contribute significantly to $\Delta G(R)$.

\subsection{Potential of mean force and hydration structure}
\label{pmf}

Figure~5 depicts $\Delta G(R)$ along the reaction coordinate
$R$.  The ZW glycine is found to be more stable
than the IIp conformer by a free energy difference, $\Delta G$, of
$11.2$ kcal/mol.  We also find a
$12.7$~kcal/mol transition state barrier, $\Delta G^*$, 
between these two forms of glycine located
at $R \approx 0.2$ \AA.  The statistical uncertainties in both $\Delta G$
and $\Delta G^*$ are estimated to be about 1.4~kcal/mol.  On the average,
this transition state structure exhibits O(1)~-~H and N~-~H distances
of $\sim 1.1$ and $\sim 1.3$ \AA,
respectively.  We will consider several corrections to $\Delta G$ and
$\Delta G^*$ below, but they will not qualitatively modify these conclusions. 

\begin{figure}
\centerline{\hbox{\epsfxsize=3.50in \epsfbox{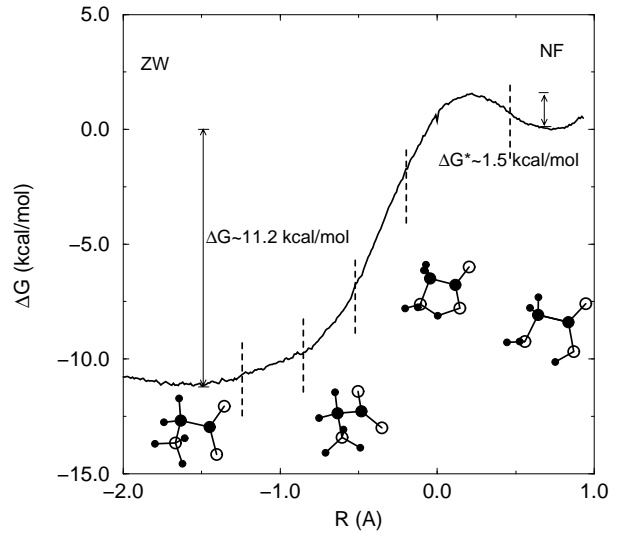}}}
\caption[]
{\label{fig5} \noindent
Potential of mean force as a function of the chosen reaction coordinate
$R=R_{\rm N-H}-R_{\rm O-H}$.  The six umbrella sampling windows
are demarcated with dashed lines, and snapshots of the glycine
molecule in some of the windows are depicted.
}
\end{figure}

Figure~6 plots as functions of $R$ the mean hydration numbers
for O(1) plus O(2), N, and the proton being transfered.  They are
compiled across all six umbrella sampling windows.  The results at the
boundary of two windows are averaged.  As $R$ increases, 
the ZW glycine continuously transforms to the IIp neutral form  and
the glycine dipole moment decreases, 
water becomes less structured around it, and the hydration
number decreases for all glycine atoms we examined.  In particular,
when $R$ reflects a IIp neutral form configuration, the NH$_2$ and
COOH groups do not strongly bind to water molecules;
each amine proton forms hydrogen bonds to only an average of $\sim 0.5$
water molecules.  As a result, the hydration numbers drop almost by a
factor of two in these cases.

\begin{figure}
\centerline{\hbox{\epsfxsize=3.50in \epsfbox{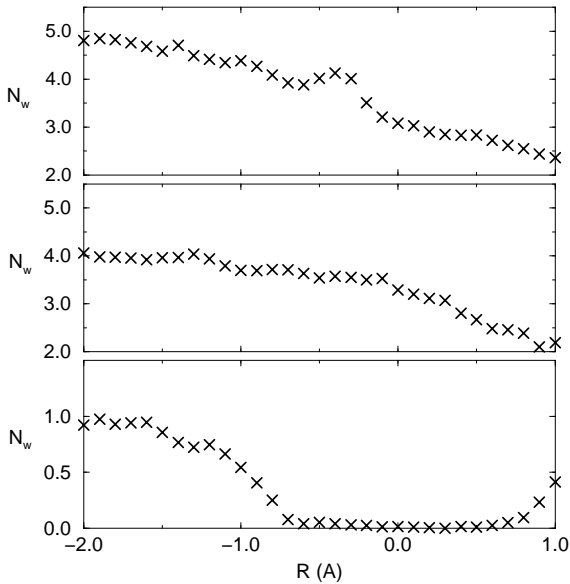}}}
\caption[]
{\label{fig6} \noindent
$R$-dependent mean number of water molecules in the first hydration
shell, N$_{\rm w}$, for various atoms.  (a) O(1) and O(2) combined;
(b) N; (c) H involved in intramolecular proton transfer.  The cut-off
criteria for determining these coordination numbers are: O~-~H$_{\rm w}$
distance of 2.5 \AA; N~-~O$_{\rm w}$ distance of 3.5 \AA; and
H~-~O$_{\rm w}$ distance of 2.4 \AA; respectively, where the subscript
``w'' refers to water molecules.  
}
\end{figure}

As for the proton being transfered, it forms covalent and intramolecular 
hydrogen bonds with N and O(1) in the range -0.8 \AA $< R < $ 0.7 \AA, 
but forms no hydrogen bonds with water molecules.
If $R$ is further increased, the intramolecular hydrogen bond cannot
be sustained, and the proton begins to establish an intermolecular
hydrogen bond with water molecules.  Thus hydrogen bond network
fluctuations are crucial to proton transfer, as observed in proton
transport in water.\cite{tuckerman02}  The statistical uncertainties
in these $R$-dependent hydration numbers are of order $\sim 0.3$.
Considering this, the hydration numbers are reasonably well converged, smooth
functions of $R$.  The predicted progression of hydration numbers should
be valuable for future quasi-chemical calculations.\cite{pratt}

At the transition state, where $R \sim 0.2$ \AA, the hydration numbers
(Fig.~6) closely resemble those found for the IIp neutral
form conformer, but differ significantly from that of the ZW.

To estimate the statistical uncertainties in our thermodynamic quantities,
we define $\Delta \Delta G = |\Delta G(R_2) - \Delta G(R_1)|$.
For windows (b)-(e), $R_2$ ($R_1$) is the maximum (minimum) $R$ used
in each window in Fig.~5.  For windows (a) and (f), $R_2$ and
$R_1$ are the $R$ values at the local minima, respectively.  The statistical
uncertainty in $\Delta \Delta G$ is estimated by dividing the 10~ps trajectory
in each window into 4 blocks of 2.5 ps each.
For windows (d) and (e), we run a trajectory twice as long.  
The estimated deviation from the mean computed using 2.5 and 5~ps blocks are
almost identical for window (d), suggesting that the correlation time for
$\Delta \Delta G$ is less than 2.5~ps.  Recall that Fig.~3(c) shows 
only 5 instances each of ``new'' water molecules entering,
and completely leaving, a hydration shell within the 10~ps trajectory segment
depicted in this window.  There are also numerous transient fluctuations
in N$_{\rm w}$, however, and these fluctuations allow statistically meaningful
sampling of $\Delta G(R)$.  The correlation time appears
slightly larger than 2.5~ps for window (e).  Given that most windows are
sampled for 10~ps, we report the cumulative uncertainty in $\Delta G$
as twice our estimated overall deviation from the mean estimated using
2.5~ps time blocks.

Recall also that windows (a) and (b) are sampled at T=350~K.  For these
two windows, we estimate $\Delta G(R)$ at T=300~K by assuming that the
free energy barrier is mainly due to entropy.\cite{slifkin}  Then
$\Delta G(R)/(k_{\rm B}T) \propto \log P(R)$,
where $P(R)$ is the probability, assumed to be temperature
independent, that the reaction
coordinate spontaneously exhibits value $R$.  As shown in Fig.~5,
$\Delta \Delta G$ in these two windows combined is 1.4~kcal/mol.
Even if $\Delta G(R)$ and $P(R)$ are entirely due to enthalpy,
the resulting small error arising from sampling these windows
at T=350~K will be a small fraction of 1.4~kcal/mol, and will have
little effect on the ZW$\rightarrow$IIp $\Delta G$.

\subsection{Neutral form glycines: spontaneous direct proton transfer}
\label{glynf}

A typical IIp neutral form glycine
 hydration structure is depicted in Fig.~2.
The carboxyl oxygen and
the two amine group protons of neutral form glycine
exhibit hydration numbers of 2.1 and 1.0, respectively. 
These numbers are obtained using the constraint of Table~\ref{table1}.

In fact, we find that a constraint is necessary to stablize the IIp
glycine molecule in AIMD and QM/MM trajectories.  Upon releasing the
constraint, the IIp molecule spontaneously undergoes a direct proton
transfer to the ZW form in a sub-picosecond time scale, regardless of
the starting configuration.  This is consistent with the small
proton transfer free energy barrier computed using umbrella
sampling.

\subsection{Comparison with experiments}
\label{comp_expt}

We predict a ZW$\rightarrow$IIp activation barrier ($\Delta G^*$) 
of 12.7 kcal/mol, which is ostensibly 
within 1.6 kcal/mol of the reported experimental $\Delta G^*$.\cite{slifkin}
We note that the experimental rate has been measured with
both nuclear magnetic resonance\cite{sheinblatt} and the thermally
modulated chemical relaxation method,\cite{slifkin} and the reported
intramolecular proton transfer rates are within 12\% of each other.
We also find a free energy difference for conversion from the ZW to 
the IIp conformer ($\Delta G$) of 11.2 kcal/mol, 
which is ostensibly 
54\% larger than the experimental value of
7.27 kcal/mol.\cite{wada,note2}  
As will be shown, various corrections to our AIMD free energies are small.

Thus we find, in qualitative agreement with quantum chemistry
calculations,\cite{bandy,tunon00}
that the experimental results do not correspond to
the neutral form IIp conformer that is the focus of this study.
The NMR result\cite{sheinblatt} states that the rate it measures
is associated with a NF product that must undergo proton exchange
with water before reverting back to the ZW form; the IIp conformer, 
which we predict to have a picosecond lifetime, will not qualify.
It is likely that the observed rate\cite{sheinblatt,slifkin}
involves the ZW first tranforming into the IIp molecule, and then on
to one or several more stable NF conformers that undergo
proton exchange in water and lead to the coalescence of NMR
line shapes.\cite{sheinblatt}  The NMR free energy
barrier will reflect a composite of these processes.

We have not computed the free energies and barriers of other
NF conformers compared to the IIp neutral form.  
From our discussion in the previous section,
our predicted $\Delta G = 11.2$~kcal/mol exceeds
the experimental value of $\sim 7$~kcal/mol, which is consistent
with the fact that IIp is only an intermediate, and a lower
free energy conformer exists.  The search for this stable conformer
using AIMD will be left to future work.

We find that $\Delta G$ predicted by AIMD and the supermolecule
approach, which uses one glycine and three water molecules 
plus dielectric continuum, differ by several kcal/mol if BLYP is
used in the latter case.  This comparison will be presented in
Appendix A along with the suggestion that the discrepancy
exists because the supermolecule lacks a fully
occupied first hydration shell.

\section{Conclusions}

Using {\it ab initio} molecular dynamics calculations and the BLYP exchange
correlation functional, we find a $12.7$~kcal/mol free energy
barrier between zwitterion and conformer IIp of the neutral form glycine
in water.  The statistical uncertainty is estimated to be of order
1.4~kcal/mol.
We predict a $11.2\pm 1.4$ kcal/mol free energy difference between
the zwitterion and this IIp conformer.  The experimental free energy
difference between the zwitterion and the neutral form is 7.27 kcal/mol,
although precisely which neutral form conformer dominates in water
has not yet been determined experimentally, and will be addressed
in future work.

We also gain useful qualitative insight on hydration structures from our
AIMD simulations.
We find that the hydration structure of the -COO$^-$ group in the
zwitterion is similar to that of the formate ion,\cite{choo_ab} forming
an average of 4.7 hydrogen bonds with water, while the -NH$_3^+$ group
forms 3.0 hydrogen bonds with water molecules, yielding 8 water molecules
in the nearest hydration shell of the ZW.  The carboxyl
oxygen and the two amine group protons of neutral form glycine
exhibit hydration numbers of 2.1 and 1.0, respectively.  The coordination
numbers along the reaction coordinate interpolate between these limits.

This work demonstrates the viability of AIMD to predict free energy
changes in aqueous reactions.  
AIMD allows fairly extensive sampling of the first hydration
shell water configurations, which is found to have strong
correlation with the progress along the reaction coordinate.
While our conclusions about the crucial role of
neutral form glycine conformers other than IIp is in agreement with
some quantum chemistry supermolecule calculations (and in disagreement
with reactive force field works\cite{yamabe1,yamabe2}), our quantitative
results may be particularly valuable toward
accurate parameterization of future quasi-chemistry calculations
of proton transfer in aqueous environments.\cite{pratt}

\section{Acknowledgement}

We thank Todd Alam and Normand Modine for useful
suggestions.  This work was
supported by the Department of Energy under Contract DE-AC04-94AL85000.
Sandia is a multiprogram laboratory operated by Sandia Corporation, a
Lockheed Martin Company, for the U.S.~Department of Energy.

\section*{Appendix A: Corrections to {\it ab initio} molecular dynamics results}

Several corrections to the thermodynamic quantities ($\Delta G$ and
$\Delta G^*$) determined from our AIMD trajectories should be considered.
We show that they will not alter our conclusions. These
include zero point energy corrections and use of a more sophisticated exchange
correlation functional.

To estimate such corrections, we conduct gas phase cluster energy
minimization calculations using
the Gaussian code, with B3LYP\cite{b3lyp_ex} and BLYP exchange correlation
functionals and various basis sets.\cite{gauss} Cluster optimizations
are carried out in the 6-31G(d) basis, to reproduce results found in the
literature, as well as in the 6-31+G(d,p) basis.  Frequency calculations
confirm that the predicted structures are true minima, and zero point
energies are computed at the same level of theory.  Refined single point
estimates of energies are obtained using an
extended 6-311+G(2d,p) basis set applied to the configurations found
in the minimization calculations.

\subsection{Zero Point Energy}

The results in Sec.~\ref{pmf} do not include zero point energy (ZPE)
corrections.  It is fairly costly to include ZPE effects in AIMD simulations
via the path integral formalism.  ZPE corrections, however, can be
estimated during post processing
of the trajectory data by appealing to supermolecule calculations.
Here we consider a supermolecule geometry similar to that in
Ref.~31.
This cluster has one glycine and three water molecules, as shown
in Fig.~7, and a polarizable dielectric continuum model\cite{bsj}
is used to treat the bulk water boundary conditions.  Using
the basis and refined basis sets described above, we find that
ZPE indeed raises the free energy of
ZW glycine by only $\sim 1$~kcal/mol relative to IIp, while the
transition state is lowered by 1.5~kcal/mol, similar
to the predictions of
Ref.~31.

\begin{figure}
\centerline{\hbox{\epsfxsize=3.50in \epsfbox{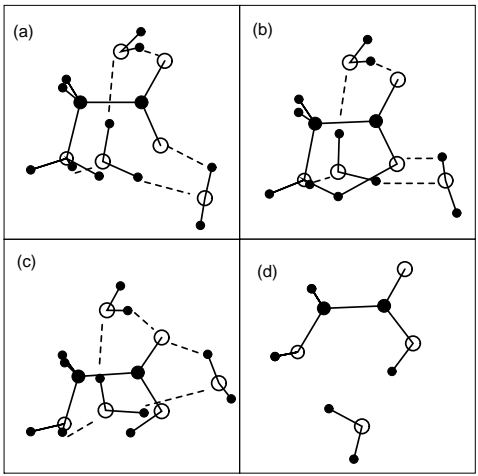}}}
\caption[]
{\label{fig7} \noindent
Structures of a gas phase cluster with one glycine plus 3 water
molecules, similar to
Ref.~31.
(a) IIp; (b) transition
state; (c) ZW; (d) bridging water.
}
\end{figure}
\subsection{Choice of exchange correlation functional}

While the BLYP exchange correlation functional used in this work reproduces
the water-water pair correlation function well,\cite{siepmann} it is
known to overestimate correlation effects in hydrogen bonded systems
and slightly underestimate proton transfer energy
barriers.  For example, from a survey of proton transfer barriers
in molecular systems,\cite{barone94,barone95,barone96,sadhuhan}
B3LYP and MP2 results are found to be within 0.6~kcal/mol
with the highest level of quantum chemistry
calculations (QCISD(T) and CCSD(T)) 
in a suite of test cases, while the BLYP $\Delta E^*$
is at most 2.3~kcal/mol less than B3LYP predictions.
This suggests that the accuracy of
the small, IIp$\rightarrow$ZW proton transfer barrier
predicted using the BLYP functional can be assessed by
comparing with B3LYP results in a suitable basis set.

Again we consider a supermolecule geometry similar to that in
Ref.~31,
using the same basis and refined basis sets as before 
The B3LYP functional predicts
the aqueous phase IIp$\rightarrow$ZW $\Delta G$ and $\Delta G^*$ are 
predicted to be -4.0~kcal/mol and 2.6~kcal/mol,
respectively.  These predictions are similar to Kassab {\it et al.}'s
$\Delta G = -5.4$~kcal/mol and $\Delta G^*= 2.2$~kcal/mol, although
those were obtained using a smaller basis set.  

The BLYP functional, which we use in our AIMD simulations,
predicts that
the aqueous phase IIp$\rightarrow$ZW $\Delta G$ and $\Delta G^*$ 
are -5.4~kcal/mol and 1.2~kcal/mol, respectively.  This confirms that 
calculations with the BLYP functional
yield similar but slightly smaller $\Delta G^*$'s than with the B3LYP
functional for the proton transfer systems of interest.  
The more reliable B3LYP model will raise $\Delta G^*$ slightly, but this
is opposite in sign to the ZPE correction and the two partially cancel.
The BLYP and B3LYP $\Delta G$'s also differ by only 1~kcal/mol.
Yet the $\Delta G $ of $-5.4$~kcal/mol predicted using the
BLYP functional with a supermolecule plus dielectric continuum
approach is considerably
smaller than the AIMD umbrella sampling prediction of
-11.2~kcal/mol, even after the latter is corrected for ZPE.

Thus there is a several kcal/mol discrepancy between our BLYP-based
AIMD results, and what we compute using BLYP and a static 3-water
supermolecule plus dielectric continuum calculation.  We tentatively
assign this to the fact that 3 water molecules are not sufficient to model 
the -NH$_3^+$ and -COO$^-$ first hydration shells, although
the supermolecule predictions also show some dependence on basis set 
and choice of dielectric continuum model.  The limitations
of using a small number of water molecules have been
pointed out already in the literature.\cite{truhlar,note5}

\section*{Appendix B: Electrostatic boundary conditions and
finite size effects in AIMD simulations}

In this section we describe the finite size corrections to 
the hydration free energy $\Delta G_{\rm hyd}$ of a glycine
zwitterion in water, which is closely related to the potential of
mean force $\Delta G (R)$ associated with the intramolecular proton
transfer in glycine.  We consider two estimates
of this correction: that due to a dielectric
continuum argument, and explicit calculations of the AIMD potential of
mean force profile by varying the simulation box size.  Both indicate that
the 11.8~\AA~simulation box is adequate for glycine zwitterion in water.

The following arguments are general and illustrate the
ability of {\it ab initio} molecular dynamics (AIMD) methods to
successfully predict the hydration free energies of dipolar species
in water.

\subsection*{Dielectric Continuum Estimate}

The effect of periodic boundary conditions on liquid state
computer simulations has been a well-studied subject.\cite{pratt_pbc}
In our case, zwitterion glycine has a relatively large dipole moment;
we compute a value of order 15 Debye in the gas phase.
Nevertheless, using a dielectric continuum estimate, we will show that
the relatively small box size has
little effect on the glycine zwitterion $\Delta G_{\rm hyd}$.

\begin{figure}
\centerline{ \hbox{\epsfxsize=1.80in \epsfbox{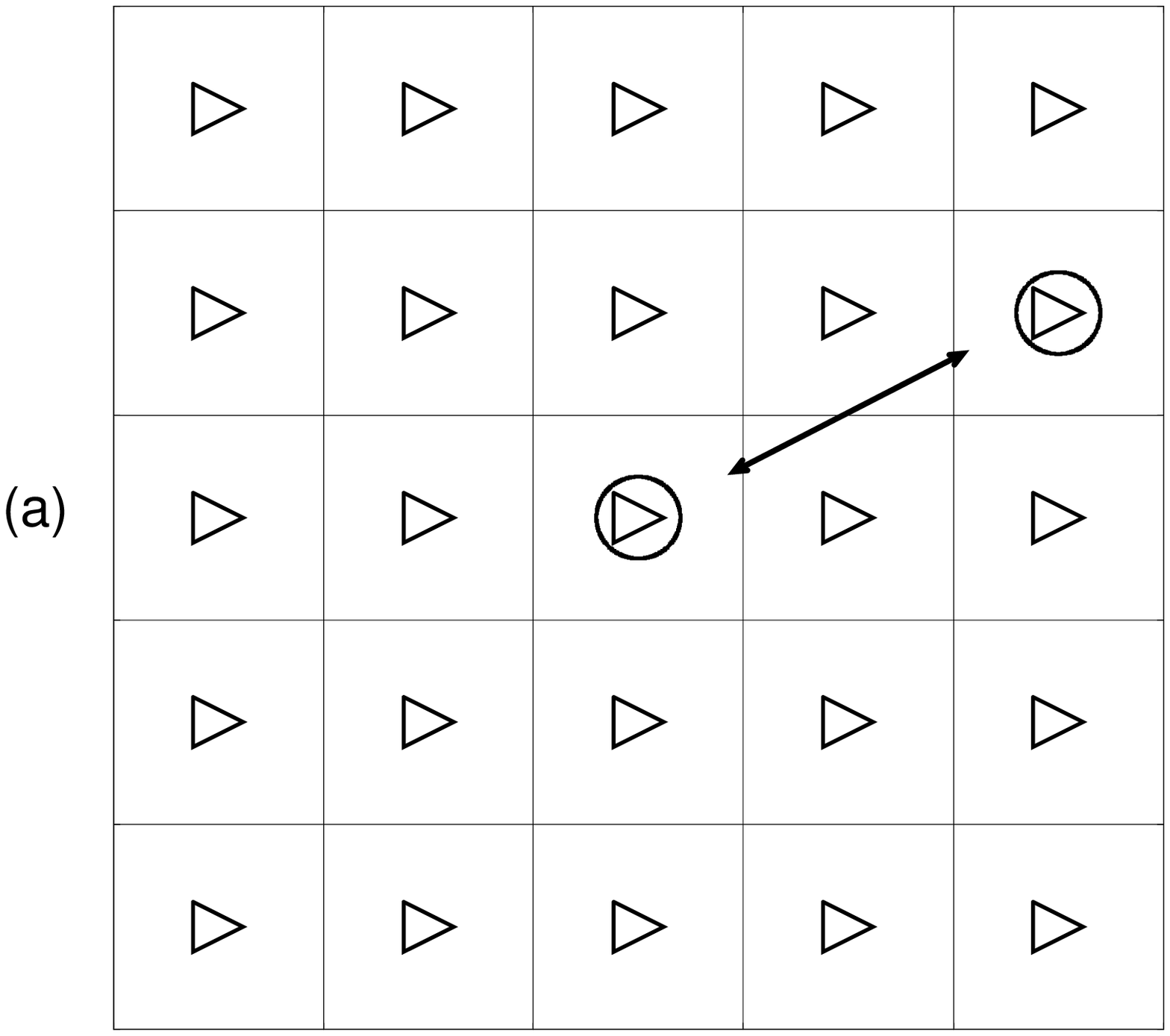}}
             \hbox{\epsfxsize=1.80in \epsfbox{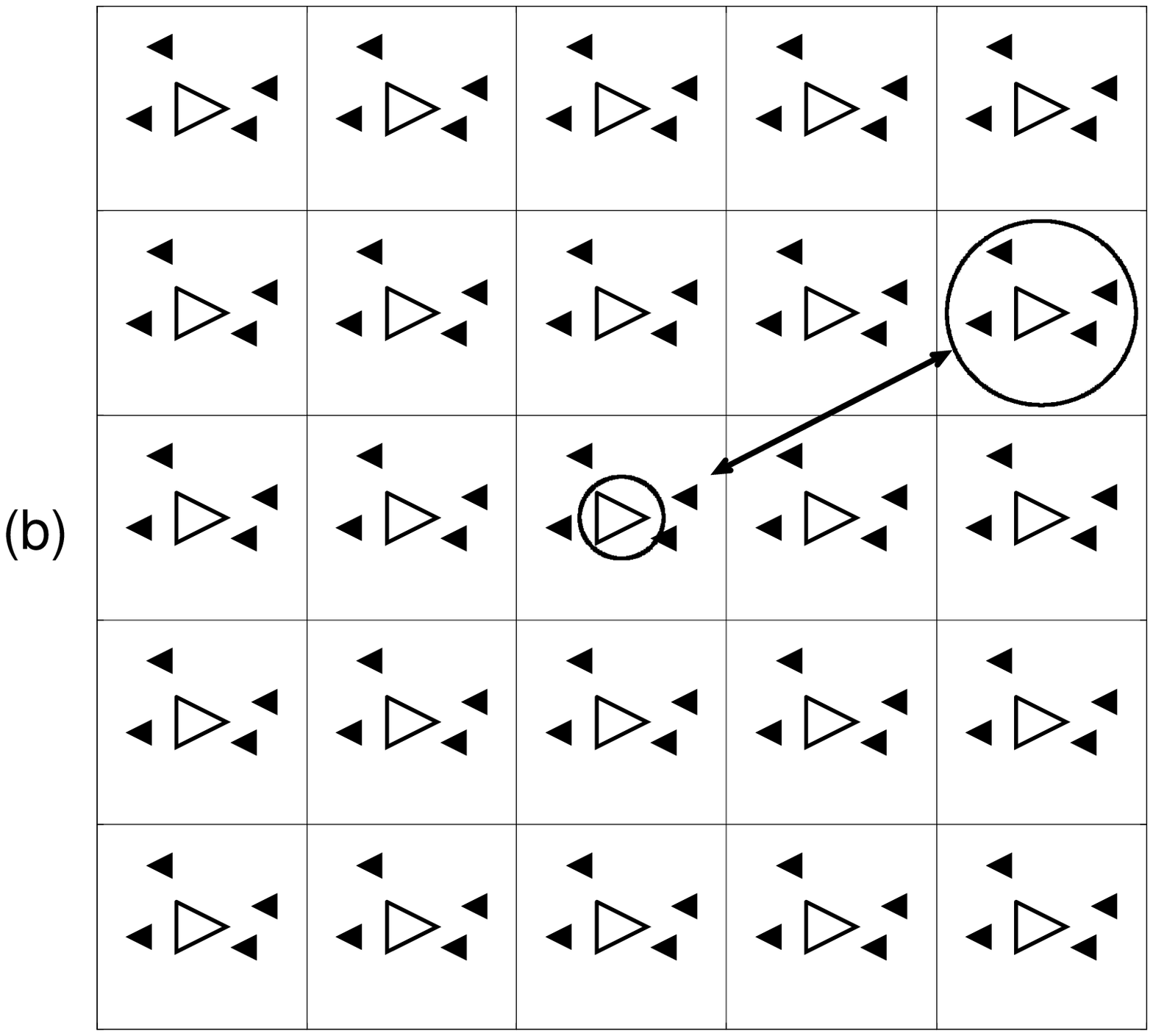}} }
\centerline{ \hbox{\epsfxsize=1.80in \epsfbox{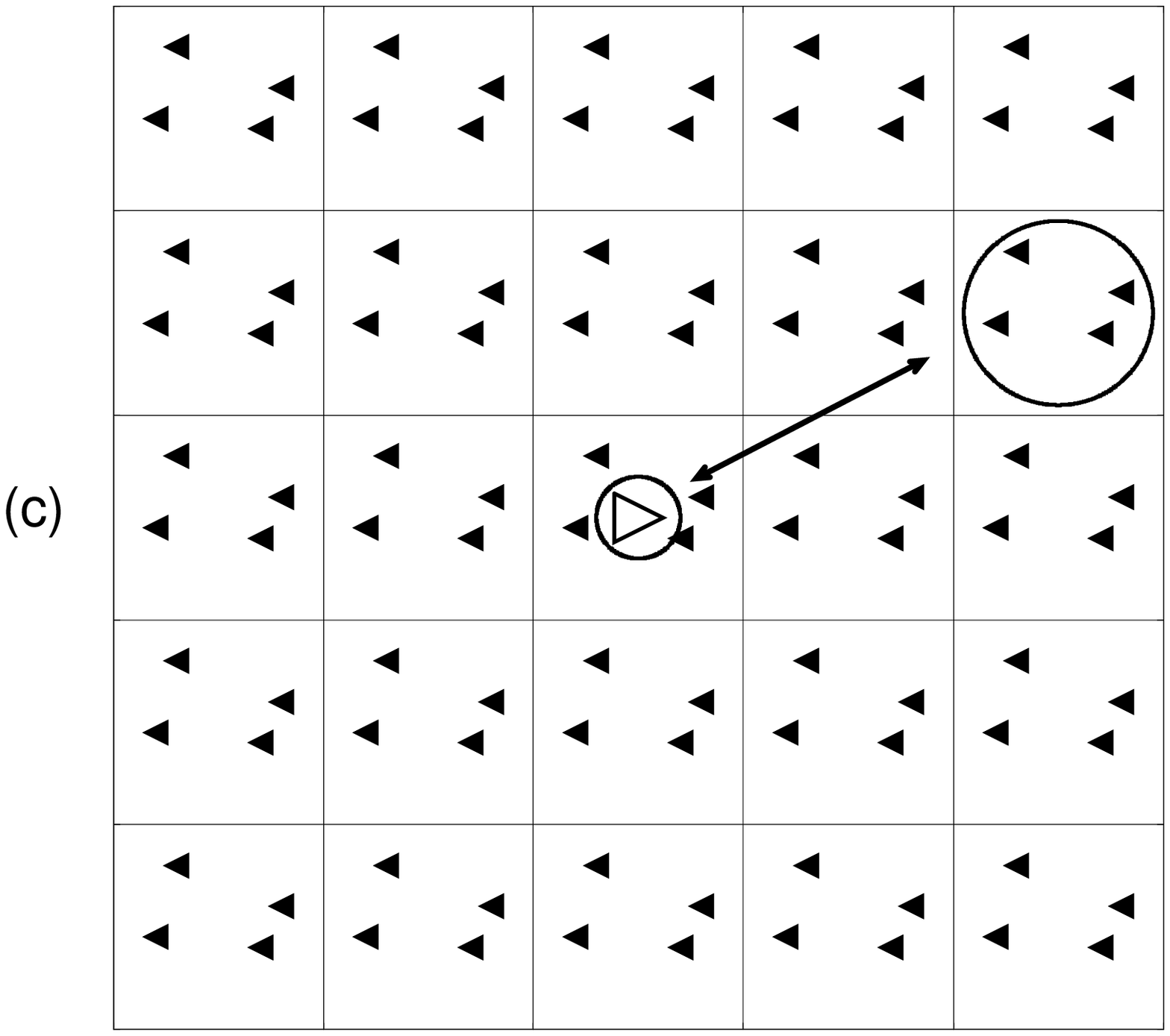}} }
\caption[]
{\label{fig9a} \noindent
Illustrations of dielectric continuum estimates for a dipolar solute
embedded in a high dielectric solvent.  Hollow triangle: solute; filled
triangle: solvent
molecules.  The triangles point in the directions of the dipole moments.
The black circles illustrate which interactions between the solute
and the periodic images are accounted for.
(a) solute-solute image interaction only (attractive,
corrected by Eq.~\ref{a1});
(b) all interactions (repulsive, corrected by Eq.~\ref{a3}); (b)
only solute-solvent
image (repulsive, corrected by Eq.~\ref{a2}).  With pair-wise additive
force fields, (a) and (c) add up to (b); thus they cancel to a
large extent in a medium with a high dielectric constant.
Despite the particulate representation of solvent depicted herein,
for the purpose of this discussion, the solvent is
treated as a dielectric continuum.
}
\end{figure}

Ewald sums and periodic boundary conditions are almost invariably used
in AIMD simulations, including those
reported in this work.  Consider the $\Delta G_{\rm hyd}$ of
a single dipolar solute (molecule) at infinite dilution computed using
Ewald summation.  In a vacuum (Fig.~8a),
the solute-solute image contribution to the computed configurational energy
is given by\cite{makov}
\begin{equation}
E_{\rm solute-solute} = - (2\pi {\bf d}^2 /3 L^3) \label{a1}
\end{equation}
in atomic units,
where ${\bf d}$ is the dipole moment and $L$ is length of the cubic
simulation cell.
This solute-solute image interaction due to the Ewald sum is
{\it attractive} in a cubic box; the boundary
conditions overestimate the magnitude of $\Delta G_{\rm hyd}$.
The correct energy of a dipolar solute in vacuum is then the value
computed using Ewald sum {\it minus} $E_{\rm solute-solute} $.

For a box size of $L=11.8$\AA\/ and $|{\bf d}|=15$ Debye, Eq.~\ref{a1}
implies a $\sim +5$ kcal/mol correction in vacuum.

Our zwitterion glycine molecule, however, resides in water, not
in vacuum (Fig.~8b).  In this case, the finite size
contribution $E_{\rm dipole}$ in Eq.~\ref{a1} is replaced by
\begin{equation}
E_{\rm screened} = - (2\pi {\bf d}^2 /3 L^3) /\epsilon_o  \label{a3}
\end{equation}
where $\epsilon_o$ is the (relative) static dielectric constant of water.
Therefore, the correct
result is the Ewald sum value plus $|E_{\rm screened}|$. A similar dielectric
continuum estimate of screened solute-solute image interaction
has been used successfully to understand the finite size effect in
the potential of mean force of sodium chloride ion pair separation,
computed using force field based molecular dynamics.\cite{haymet}
In that case, as the sodium and chloride ions are pulled apart from
their contact ion-pair configuration, a large dipole moment is
incurred.  Nevertheless, due to the strong aqueous dielectric screening,
the use of Ewald sum and periodic boundary conditions still lead to
results well converged with box size (see Figs.~1~\&~2 of
Ref.~77.)
Estimates similar to
Eq.~\ref{a3} are also used in solid state density functional
theory calculations of defect energetics.\cite{makov}

Assuming AIMD water exhibits $\epsilon_o \sim 80$,
the dielectric continuum estimate of
Eq.~\ref{a3} suggests that the Ewald sum we use
entails a correction of less
than $+0.1$ kcal/mol to $\Delta G_{\rm hyd}$ for
our box size of $L=11.8$~\AA.

In hydration free energy calculations that employ classical force
fields, the corrections due to Ewald sums and periodic boundary
conditions are often described from a different perspective.\cite{pratt95}
Unlike the case with density functional theory, where the
interaction is inherently many-body and not pairwise
decomposible, classical force fields typically allow the computation
of strictly solute-water interactions.  Thus the interaction of
a dipolar solute with {\it all} periodic images, Fig.~8b,
can be unambiguously separated into solute-solute image (Fig.~8a)
and solute-water image (Fig.~8c) contributions.

Using explicit force field-based molecular dynamics simulations, Hummer,
Pratt, and Garcia\cite{pratt95} have eloquently discussed the
ramification of this separation.  If only the solute-water terms is used,
the long range solute-{\it water image} interaction,
pictorially depicted in Fig.~8c, can lead to considerable finite
size dependence in $\Delta G_{\rm hyd}$.  If the entire expression
of Fig.~8b is used, i.e., if solute-solute self-energy term
(Fig.~8a) is also included, finite size dependences become
negligable.\cite{pratt95} From such an analysis, it can be deduced that
the solute-water image interaction is {\it repulsive}:
\begin{equation}
E_{\rm solute-water} \approx +(2\pi {\bf d}^2 /3 L^3),  \label{a2}
\end{equation}
Equation~\ref{a2} {\it decreases}  $\Delta G_{\rm hyd}$.
It is equal and opposite to Eq.~\ref{a1}, the self-energy Makov and Payne
corrected for a dipole in vacuum, as it should be;
compare the expressions in Eq.~19,
Ref.~78,
with the last term of Eq.~6 in
Ref.~76.

The physics behind the success of the solute-solute image self-term
in removing finite size effects from the solute-water image interaction
(Fig.~8c) is as follows.
The solute with dipole moment ${\bf d}$ causes water 
dipole moments in its vicinity to align against it.  Since water
has a large dielectric constant, to a zeroth approximation
the screening of the solute dipole moment is complete, and
water molecules proximal to the solute exert a $\sim -{\bf d}$ dipole
moment per simulation cell.  (On average, the simulation cell should have
a net dipole moment that approaches zero.  For an anology, consider
a metal, which has an infinite dielectric constant and cannot support
internal electric fields; in that case, it is obvious each unit cell
has a zero dipole moment, as long as localized Wannier functions
are judiciously used to demarcate the centers of charge.)  Thus the
solute-water images in the periodically replicated system give a
{\it repulsive} energy of $+(2\pi {\bf d}^2 /3 L^3)$
(Fig.~8b), equal and opposite to the solute-solute
image interaction.

The solute-solute image and the solute-water image interactions
are not pairwise additive in AIMD simulations, where energies are
inherently many-body in nature.  Instead, AIMD propagates Newton's
equations of motion according to the total energy and forces computed
using Ewald summation, which automatically include both long-range
solute-solute image (Fig.~8a) and solute-water image
interactions (Fig.~8c) .
Thus AIMD simulations do {\it not} suffer from the large finite
size correction associated with Eq.~\ref{a2}, namely the unfavorable
solute-water image interaction (Fig.~8c) --- as long as $\epsilon_o$
is large and the dielectric continuum approximation is valid.

Since water does not have an infinite dielectric constant,
the screening of the solute dipole is incomplete, and the solute-water
image repulsion (Fig.~8c) should be slightly less than the
solute-solute image attraction (Fig.~8a).
Equation~\ref{a3} is an estimate of this residual effect.
In the $\epsilon_o \rightarrow \infty$ limit, the system is metallic
and this correction vanishes.  When $\epsilon_o=1$, the system
is in a vacuum, and Eq.~\ref{a3} correctly reduces to Eq.~\ref{a1}.
For $\epsilon_o \approx 2$, the above arguments may not be valid
because they rely on the assumption of strong dielectric screening.

In summary, our dielectric continuum estimate indicates that
the finite size correction for our AIMD glycine zwitterion hydration
free energy is less than 0.1~kcal/mol.

\subsection*{Explicit Simulation Results}

The above analysis assumes a dielectric continuum description
of water.  The particulate nature of water is not taken into account;
with a relatively small simulation cell size and
a small number of water molecules, it may not be completely valid.
To investigate this, we have explicitly varied the simulation
cell size in AIMD calculations of the glycine intramolecular
proton transfer potential of mean force ($\Delta G(R)$)
in two umbrella sampling windows.  We use a simulation cell
roughly twice the volume used in the main text, containing
98 instead of 52 water molecules.  10~ps AIMD trajectories
are used for collecting statistics.

\begin{figure}
\centerline{ \hbox{\epsfxsize=3.50in \epsfbox{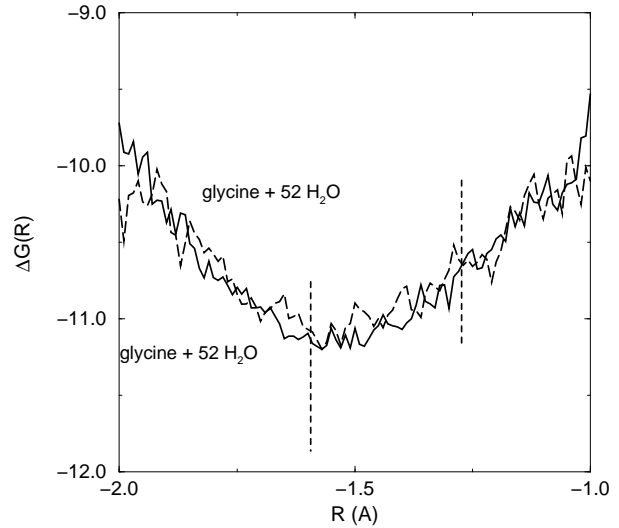}}}
\caption[]
{\label{fig8} \noindent
Finite size effect in $\Delta G(R)$ for glycine in sampling window
(f).  The blue lines demarcate the $R$ range used in this window
for the overall potential of mean force.  This is the unconstrained
zwitterion window; a wide range of $R$ values are sampled,
and the statistics are not as good as in other windows.
}
\end{figure}

First we consider the unconstrained zwitterion sampling
window (a).  See Fig.~9.  
This sampling window does not use constraints, and so the
glycine zwitterion conformations fluctuate more than in other windows.
As a result, the statistics for $\Delta G(R)$ tend to be worse
than for other windows.  Nevertheless, it can be seen from Fig.~9
that $\Delta G(R)$ predicted for the two simulation cell sizes are
within statistical uncertainties of each other within a range
of $R$ where $\Delta G(R)$ does not vary by more than 3 k$_{\rm B}$ T.
The mean hydration structures are similar in both trajectories as well.
The average hydration numbers of the larger and smaller simulation cells
are with 0.1 water molecules of each other, with the former
having 20\% more intramolecular hydrogen bond --- well within
the statistical uncertainty.  

\begin{figure}
\centerline{ \hbox{\epsfxsize=3.50in \epsfbox{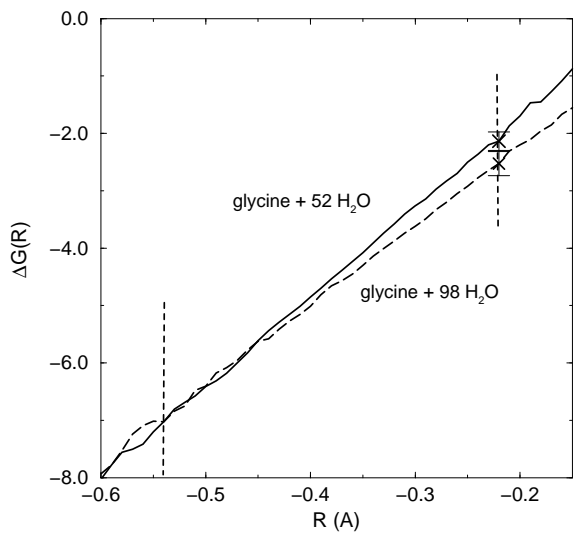}}}
\caption[]
{\label{fig9} \noindent
Finite size effect in $\Delta G(R)$ for glycine in sampling window
(d).  The blue lines demarcate the $R$ range used in this window
for the overall potential of mean force.
}
\end{figure}

Since the $\Delta G(R)$ profile in window (a) is relatively flat,
we also look at window (d).  This window contributes
half ($\sim 5$ kcal/mol) of the overall $\Delta G(R)$, and should
shed more light on finite size effects.  In Fig.~10,
we see that the larger and smaller simulation cells yield
$\Delta \Delta G$ of 4.5 and 5 kcal/mol, respectively.
The difference is within the statistical uncertainty.  From these
explicit simulation results, we conclude that the finite size
effect should be within our statistical uncertainty.

\newpage



\begin{references}
\bibitem{helmholtz}
Our fixed volume predictions actually correspond to the Helmholtz
free energy.  Nevertheless, we will use ``$\Delta G$'' throughout,
since that appears to be the common usage in this field.
\bibitem{h_num}
The average hydration number, N$_{\rm w}$, obtains by integrating the pair
correlation $g(r)$ of the pertinent atom with the water O or H atom
to its first minimum.
\bibitem{note2}
Recall that a QM/MM estimate is even
larger.\cite{shoeib}
\bibitem{note5}
Note that an MP2 based, 6 water supermolecule plus dielectric
continuum calculation predicts $\Delta G = 10.3$ kcal/mol after
correcting for ZPE.\cite{fern}  However, this supermolecule
has a bridging water molecule between the -NH$_3^+$ and -COO$^-$
groups, which is not observed in our simulations.
\end{references}
\end{document}